\documentstyle[draft]{mn}
\input epsf

\def\epsn@scaling{1.0}
\def\epsnscale#1{\gdef\epsn@scaling{#1}}
\def\plotone#1{\centering \leavevmode
\epsfxsize=\epsn@scaling\columnwidth \epsfbox{#1}}

\def\eepsn@scaling{1.65}
\def\eepsnscale#1{\gdef\eepsn@scaling{#1}}
\def\pplotone#1{\centering \leavevmode
\epsfxsize=\eepsn@scaling\columnwidth \epsfbox{#1}}

\def\eeepsn@scaling{0.80}
\def\eeepsnscale#1{\gdef\eeepsn@scaling{#1}}
\def\ppplotone#1{\centering \leavevmode
\epsfxsize=1.70\columnwidth \epsfysize=1.45\columnwidth \epsfbox{#1}}

\def\eeps@scaling{1.5}
\def\epsscale#1{\gdef\eeps@scaling{#1}}
\def\plotonen#1{\centering \leavevmode
\epsfxsize=\eepsn@scaling\columnwidth \epsfbox{#1}}

\begin{document}

\title[Monthly Notices: Sinking Satellites and the Heating of Galaxy Discs]{Sinking Satellites and the Heating of Galaxy Discs}

\author[ Vel\'azquez and White ]
{ H\'ector Vel\'azquez$^{1,2,3}$ and Simon D.M.
White$^2$ \\
$^1$ Institute of Astronomy, Madingley Road, Cambridge CB3 0HA, UK\\
$^2$ Max-Planck-Institut f\"ur Astrophysik, Karl-Schwarzschild-Strasse
1, D-85740 Garching bei M\"unchen, Germany\\
$^3$ Observatorio Astron\'omico Nacional-UNAM, Apdo. Postal 877, 22800 Ensenada, B.C., M\'exico
}
\maketitle

\begin{abstract}

We have carried out a set of self-consistent N-body simulations
to study the interaction between disc galaxies and merging satellites
with the aim of determining the disc kinematical changes induced by 
such events. We explore a region of the parameter space embracing
satellites with different masses and internal structure and orbits of various
eccentricities. We find that the analytic estimates of T\'oth and
Ostriker (1992) are high; overestimating the disc heating 
and thickening resulting from the accretion process by a factor of about 
$2-3$. We find that the heating and thickening of the disc differ for 
satellites on prograde and retrograde orbits. The former tend to heat 
the stellar disc while the
latter primarily produce a coherent tilt. For instance, a
satellite of a Milky Way type galaxy with an initial mass of $20$\%
that of the disc and on a retrograde orbit increases the velocity
ellipsoid at the Solar neighborhood by $(\Delta \sigma_R, \Delta
\sigma_\phi, \Delta \sigma_z)_\odot \approx (11,9,6)$ kms$^{-1}$ and
produces a maximum 
increment of the vertical scale length and the stability parameter
$Q$ inside the solar radius of $300$ pc and $0.8$, respectively,
increases of about $43$\% and $53$\%. The same satellite
but on a prograde orbit
leads to changes of $( \Delta \sigma_R, \Delta \sigma_\phi, \Delta
\sigma_z)_\odot \approx (22,15,12)$ kms$^{-1}$, $\Delta z_{o,\odot}
\approx 550$ pc and $\Delta Q_\odot \approx 1.2$. Thus, disc galaxies may
accrete quite massive satellites without destroying the disc,
particularly, if the orbits are retrograde. We also find that a massive bulge  
may play a role in reducing these effects. We have quantified the
importance of the responsiveness of the halo by
replacing it by a rigid potential in several simulations. In
these cases, the increase of the vertical scale length is larger by a
factor of $1.5-2$, indicating that a self-consistent treatment is
essential to get realistic results. A frequent by-product of the
accretion process is
the formation of weak stellar warps and asymmetric discs. Finally, we
have checked how well Chandrasekhar's dynamical friction formula
reproduces the sinking rates in several of our experiments. We find
that it works well provided a suitable value is chosen for the Coulomb
logarithm and the satellite mass is taken to be the mass still bound
to the satellite at each moment.

\end{abstract}

\nokeywords
\section{Introduction}

Most currently popular models for the formation of galaxies and larger
structures postulate that growth occurs hierarchically through
gravitational clustering in such a way that small objects form first 
and then aggregate into larger systems (Frenk et al. 1988, Carlberg
and Couchman 1989, Kauffman and White 1993, Lacey and Cole 1993). The 
observational evidence seems to support these models since it is
possible to trace the influence of merging both on galaxies and on
clusters of galaxies. We can classify mergers roughly according to the
masses of the objects involved: (1) major mergers and (2) minor
mergers. The first involve galaxies of comparable mass and are often  
invoked as a mechanism to form elliptical galaxies
from spiral systems (Toomre 1977, Negroponte and White 1983, Schweizer
1990, Barnes 1992, Silk and Wyse 1993). Minor mergers involve a giant 
galaxy accreting a small satellite. The Large Magellanic Cloud and our
own Galaxy are a clear example of such an ongoing minor merger. In
spite of a general consensus that minor mergers may drive internal 
evolution in galaxies, the consequences of these events are still not
clear (Quinn and Goodman 1986, Quinn, Hernquist and Fullagar
1993, Walker, Mihos and Hernquist 1994 and, Huang and Carlberg 1997; 
hereafter QG, QHF, WMH and HC, respectively).\\

An important constraint on the effects of minor mergers was raised
by Ostriker (1990) and T\'oth and Ostriker (1992, hereafter TO).
In a high density Universe dominated by cold dark matter (CDM) about 
$80$ per cent of the dark haloes have undergone a merger in the past $5$
billion years which increased their mass by 10  per cent or more
(Frenk et al. 1988, Kauffman and White 1993, Navarro, Frenk and White 
1994). TO argued that such a merger rate is too high to be compatible
with the observed thinness and coldness of discs in spiral galaxies. 
Using a semi-analytic treatment of the problem, they derived an 
uncomfortably low upper limit for the mass that could be accreted by 
a disc like that of the Milky Way -- no more than $4$ per cent of the 
present mass within the solar circle could have been accreted during 
the last $5$ billion years, given the observed local values of Toomre's 
stability Q-parameter and of disc scale-height. They argued that this
constraint favours a low density universe (perhaps with a cosmological
constant) in which the expected merger rate is low.\\

Several complementary lines of investigation have been used to
evaluate the argument proposed by TO. One approach
concentrates on understanding the later stages of merging and the
dynamical evolution of disc structure.
Semi-restricted and full N-body simulations
have been employed to explore how infalling satellites perturb 
discs (QG, TO, QHF, WMH, HC). The most recent work by WMH and
HC used full N-body simulations and incorporated a key
ingredient in the accretion process, the responsiveness of the
halo. A possible weakness of all these studies is that they 
focussed on satellites on nearly
circular orbits. For instance, WMH  followed their satellite only
after it was already within $21$ kpc of the galaxy centre; this choice
was imposed by their decision to use a very large number of particles 
(500,000) in order to reduce numerical noise and to delay the
growth of a bar in their model disc. They found that accretion of a 
satellite with $10$ per cent of the disc mass was already enough
to provoke a 60 percent thickening of the stellar disc 
at the solar circle. In contrast, HC found that satellites with
masses between $10$-$30$ per cent of the disc mass could be put on
near-circular orbits at about $10$ disc half-mass radii,
and would be sufficiently disrupted by tides before interacting
strongly with the disc that their effects on it are quite small.
The present paper is a continuation and extension of this line of
research.\\

A complementary approach concentrates on evaluating the rate at which
satellites are accreted as a function of mass and orbital parameters.
Observation-based arguments can be used to
estimate the current merger rate either for dark haloes or for 
galaxies. Thus, signs of disturbance like tidal tails and shells 
can be considered signs of a recent merger and thus allow the merger 
rate of luminous galaxies to be obtained. From a sample of
$4000$ galaxies, 10 were identified by Toomre (1977) as results of a
recent merger. From this he was able to derive a lower limit of
$0.005$ Gyr$^{-1}$ for the current merger rate. A higher merger rate 
of about $0.04$ Gyr$^{-1}$ was found by Carlberg, Pritchet \& Infante
(1994) based on the observed numbers of close galaxy pairs and
the assumption is that pairs will merge if their closest approach
distance and relative velocity are less than their characteristic 
radius and their internal velocity dispersion, respectively (Aarseth
and Fall 1980). On larger scales it is clear that substructure is a 
common feature of clusters of galaxies, suggesting that many of them 
have formed recently by the merging of several smaller systems
(Dressler and Shectman 1988; Jones and Forman 1992; Richstone, Loeb 
and Turner 1992). Using another version of the TO argument, the
observed amount of substructure in clusters can give an estimate of
the current merging rate and so of the cosmic density parameter
$\Omega$ (Richstone, Loeb and Turner 1992; Lacey and Cole 1993;
Kauffman and White 1993; Evrard et al. 1994). Neither of these
arguments, however, can give the satellite accretion rates needed to
evaluate the TO argument.\\

Satellite accretion rates can be estimated from theoretical arguments
based either on the Press-Schechter model for hierarchical clustering
(Lacey and Cole 1993) or on high resolution simulations. For example,
Navarro, Frenk and White (1994, 1995) have carried out a series of
simulations of the formation of galaxy-satellite systems in an 
$\Omega = 1$ CDM universe; their results suggest that the existence of
thin discs may, perhaps, be reconciled with such a model, because
discs are less efficient at accreting material than are their
surrounding dark haloes. In these simulations fewer than 30 per cent
of the discs grew by more than 10 per cent in mass over the last 5 
Gyr, whereas about $80$ of the haloes grew by this much or more.
The distributions of orbital orientation and orbital eccentricity of
their satellites were essentially uniform. Thus,
Lacey and Cole's (1993) suggestion that the TO constraint might be
avoided if satellites are primarily on near-circular orbits does not
seem to be viable in a realistic hierarchical clustering model. Note
that while the simulations of Navarro et al. (1995) give useful 
indications about the rates
of satellite accretion onto discs, their resolution is too low to
give information about how this accretion affects disc structure.\\

In the present paper we address the heating of the
stellar disc by infalling satellites following the first approach 
discussed above. We consider satellites with a variety of internal 
structures and on orbits with a variety of initial orientations and
eccentricities. We follow the evolution of the systems by using
full N-body simulations of all the components. Our paper is organized as
follows: section 2 contains brief descriptions of the models we adopt
for our primary and satellite galaxies, of our numerical methods, and
of the parameters of the set of simulations we have carried out. 
Section 3 studies how the disc is heated and thickened by
infalling satellites and a comparison with TO's results is given in
section 4. We discuss the tilting and warping of the disc resulting
from such events in section 5. Section 6 concentrates on the
disruption of the satellite and the evolution of its orbit. We show,
in section 7, that the latter can be well reproduced by Chandrasekhar's local
formulation of dynamical friction. In section 8 we replace our
``live'' halo by a rigid one to demonstrate how halo response affects
the accretion and disc heating processes. Finally, in section 9, we
summarize our main conclusions.\\


\begin{table}
\begin{center}
\centering
\label {symbol1}
\caption{ Galactic parameters.}
\begin{tabular}{|l|c|l|} \hline 
      	&  Symbol &  Value \\ \hline \hline
Disc: 	&            &              			 	\\
	&  $N_D$     & $40\,960$					\\
      	&  $M_D$     & $ 5.6 \times 10^{10}$ M$_\odot$ 		\\
      	&  $R_D$     & $ 3.5$ kpc \\
	&  $z_o$     & $ 700 $ pc\\
	&  $Q_\odot $ & $1.5$ 					\\
	&  $R_\odot $ & $8.5$ kpc \\
	&$\epsilon_D$& $175$ pc \\ \hline
Bulge:  &	     &						\\
	&  $N_B$     & $4\,096$					\\
	&  $M_B$     & $1.87 \times 10^{10}$ M$_\odot$		\\
	&  $a^*$     & $525$ pc		\\
	&$\epsilon_B$& $175$ pc \\ \hline
Halo:	& 	     & 						\\
	& $N_H$	     & $171\,752$					\\
	& $M_H$      & $7.84 \times 10^{11}$ M$_\odot$ \\
	& $\gamma$   & $3.5$ kpc \\
	& $r_{cut}$  & $84$ kpc \\
	&$\epsilon_H$& $175$ pc \\ \hline
\end{tabular}
\end{center}
\medskip
$^*$ The bulge half-mass radius is $1.27$ kpc. $N_D$, $N_B$ and $N_H$
correspond to the number of particles used for each galaxy component. 
The system of units is such that $G=M_D=R_D=1$.

\end{table}

\section{Numerical Preliminaries}

In this section we describe briefly the idealised models which we
adopt to describe the primary disc galaxy and the satellites which
merge with it. We also describe the numerical tools employed to
follow the dynamical evolution of the primary/satellite system, and
the parameters which define the specific set of simulations which we
have carried out.

\subsection{The Primary Galaxy Model}

We use the methods of Hernquist (1993) to set up a self-consistent 
N-body realization of a galaxy model consisting of three
components: a disc, a bulge and a halo.  A detailed description of 
the technique can be found in Hernquist's paper. The density
distributions of the three components are: \\ 

\begin{equation}
\rho_D(R,z)\,=\,{{M_D}\over{4 \pi R_D^2 z_o}} \exp(-R/R_D)
\hbox{ sech}^2(z/z_o),
\end{equation}

\begin{equation}
\rho_B(r)\,=\,{{M_B}\over{2 \pi}}{{a}\over{r(a+r)^3}},
\end{equation}

\begin{equation}
\rho_H(r)\,=\,{{M_H \alpha}\over{2 \pi^{3/2}r_{cut}}}{{\exp
(-r^2/r_{cut}^2)}\over{r^2+\gamma^2}}.
\end{equation}

Here, $M_D$, $M_B$ and $M_H$ correspond to the masses of the disc, the
bulge and the halo, repectively. $R_D$ and $z_o$ are the radial and vertical
scale lengths of the disc. $a$ defines the scale length of the bulge
and corresponds to a half-mass radius of $a(1+\sqrt{2})$ (Hernquist
1990). Finally, $\gamma$ and $r_{cut}$ are the core and cut-off radii
for the halo and $\alpha$ is a normalisation constant. Notice that we
assume both the bulge and the halo to be spherical.\\ 

The velocities are derived from the Jeans equations (e.g. Binney and
Tremaine 1987). Isotropic gaussians are assumed for the halo and
bulge velocity distributions. For the disc, the square of the radial
velocity dispersion is taken to be proportional to the surface density
of the disc, $\sigma_R^2 \propto \exp(- R_D/R)$ (Lewis and Freeman
1989) and the vertical component of the velocity ellipsoid is
determined from $\sigma_z^2=\pi G \Sigma (R) z_o$ in agreement with an
isothermal sheet (Spitzer 1942). The azimuthal component is obtained
from the epicyclic approximation, $\sigma_\phi^2=\sigma_R^2
\kappa^2/(4 \Omega^2)$. Finally, the constant of proportionality is
determined by fixing the value of Toomre's stability Q-parameter to a
given value. We select $Q_\odot = Q( R_\odot) =1.5$ at the Solar radius. \\

We have chosen a system of units such that $U_m=M_D=1$, $U_l=R_D=1$ and
$G=1$. For a disc mass of $5.6 \times 10^{10}$ M$_\odot$ and a disc radial 
scale length of $3.5$ kpc (Bahcall, Schmidt and Soneira 1983) the units of 
time and velocity are $1.3 \times 10^7$ yr and $262$ kms$^{-1}$, 
respectively.  The half-mass radius of the disc is $\sim 1.7\,R_D$ with 
a rotation period at this radius of about $13$ time units. The model can 
be easily scaled through the following expressions for the time and velocity 
units:

\begin{equation}
U_t\,=\,4.709\times 10^{11}\,\Big{(}{U_l^3\over U_m}\Big{)}^{1/2}\, \hbox{yr}
\end{equation}
 
\begin{equation}
U_v\,=\,2.076 \times 10^{-3}\,\Big{(}{U_m\over U_l}\Big{)}^{1/2}\,\hbox{kms}^{-1}
\end{equation}
 
\noindent where $U_m$ and $U_l$ are given in solar masses and 
kpc, respectively. In Table 1 we summarize the values of the 
parameters that define our primary galaxy model. \\

We should notice that 
(1) our halo is probably too small to be realistic.
Studies of satellites in the Local Group and around external
galaxies show that galactic haloes extend to radii beyond $200$ kpc
with masses exceeding $2 \times 10^{12}$ M$_\odot$ (Zaritsky et al.
1989, Zaritsky and White 1994). However, our halo is consistent with
the largest velocities observed for halo stars in the solar neighborhood
(Carney and Lathman 1987) and should be massive enough to give
realistic orbital velocities for eccentric satellite orbits. (2) Our 
halo is maybe too concentrated. Persic, Salucci and Stel (1996) argue for a 
halo core radius of about $1-2\,R_{opt}$ where $R_{opt}=3.2\,R_D$ is the 
optical radius. However, a model with such a halo will be prone to 
form a bar which is an undesirable additional source of disc heating and 
to prevent its growth it will be necessary to increase the bulge 
contribution. The values of the parameters listed in Table 1 guarantee 
stability against bar formation of the disc galaxy model in isolation (Vel\'azquez 
and White, in preparation). \\
 
The rotation curve of our galaxy model (solid lines) is shown in figure 1. 
For comparison, two other rotation curves are displayed for different bulge 
masses but with the same disc and halo. The rotation curve for our galaxy 
model at the optical radius is $V_C(R_{opt})\approx 243$ kms$^{-1}$ in model units while the 
models with a bulge mass of $M_B=0.2\,M_D\,(=1.12 \times 10^{10}\,\hbox{M}_\odot)$ and $M_B=2/3\,M_D\,
(=3.73 \times 10^{10}\,\hbox{M}_\odot)$ have values of 
$V_C(R_{opt})\approx 236$ kms$^{-1}$ and $V_C(R_{opt})\approx 257$ kms$^{-1}$, respectively. We can 
observe that, at difference of the sample of rotation curves given by 
Persic et al. (1996), the rotation curve of our model shows a steeper rise 
in the inner region which is consistent with the observed nuclear rotation curves of galaxies in the CO-line emission and with the optical rotation 
curves for large bright galaxies (Sofue et al. 1997, Courteau 1997).  Furthermore, it peaks near $R_{opt}$ in agreement with the results of 
Courteau (1997). \\


\begin{figure}
\plotone{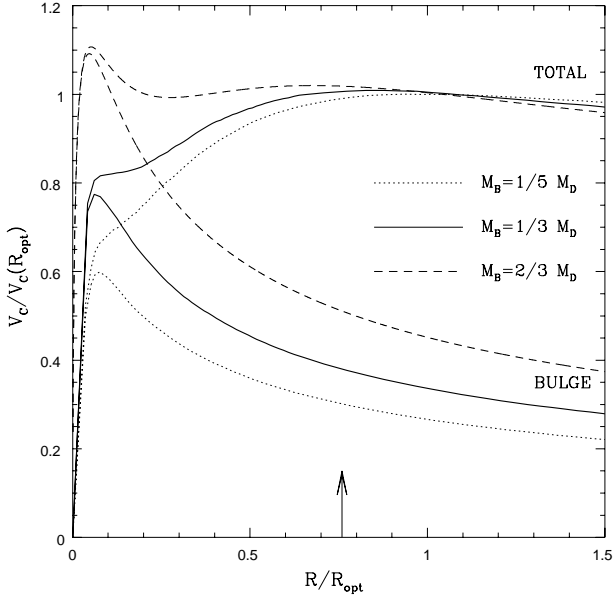}
\caption{ The rotation curve for our disc galaxy model 
(solid lines) has a value of $V_C(R_{opt})\approx 243$ kms$^{-1}$ where 
$R_{opt}=3.2\,R_D$. Dotted-lines and dashed-lines 
correspond to rotation curves for bulge masses of $0.2\,M_D$ and $2/3\,M_D$ 
with values of $V_C(R_{opt})\approx 236$ kms$^{-1}$ and $V_C(R_{opt})\approx 257$ kms$^{-1}$, 
respectively. For clarity the contribution of the disc and halo are not 
shown. The arrow indicates the Solar position.}
\label{Fig. 1}
\end{figure}

Finally, we should mention that for this galaxy model we have used a 
total number of  $216\,808$ particles. Most of these are in the halo 
component $(171\,752)$ where large numbers are required to reduce 
heating of the disc by two-body relaxation effects. \\


\begin{table}
\begin{center}
\centering
\label {symbol2}
\caption{ Satellite models.}
\begin{tabular}{|l|c|l|} \hline
 Model  &  Symbol    & Value \\ \hline \hline
 S1:	&            &              			 	\\
      	&  $M_S$     & $ 5.60 \times 10^9$ M$_\odot$ 		\\
      	&  $r_c$     & $1$ kpc \\
	&  $c$       & $0.8$ \\  
	&  $\rho_c$  & $0.52$ M$_\odot$/pc$^3$ \\ 
	&  $\sigma_c$& $52$ kms$^{-1}$	\\	\hline
 S2:    &	     &						\\
      	&  $M_S$     & $ 5.60 \times 10^9$ M$_\odot$ 		\\
      	&  $r_c$     & $ 500$ pc \\
	&  $c$       & $ 1.1$				    	\\ 
	&  $\rho_c$  & $0.84$ M$_\odot$/pc$^3$ \\ 
	&  $\sigma_c$& $60$ kms$^{-1}$ \\ \hline
 S3:    &	     &						\\
      	&  $M_S$     & $ 1.12 \times 10^{10}$ M$_\odot$	\\
      	&  $r_c$     & $ 875$ pc \\
	&  $c$       & $ 1$				    	\\ 
	&  $\rho_c$  & $ 1.36 $ M$_\odot$/pc$^3$ \\ 
	&  $\sigma_c$& $ 71$ kms$^{-1}$ \\		\hline
\end{tabular}
\end{center}
\medskip
$c$ defines the concentration of the satellite model and is given by
$\log_{10}(r_{t}/r_c)$ where $r_c$ and $r_t$ are, respectively, the
core and tidal radii of a King model. $M_S$, $\rho_c$ and $\sigma_c$ denote 
the satellite mass, the central density and the central one-dimensional 
velocity dispersion, respectively.
\end{table}


\begin{figure}
\plotone{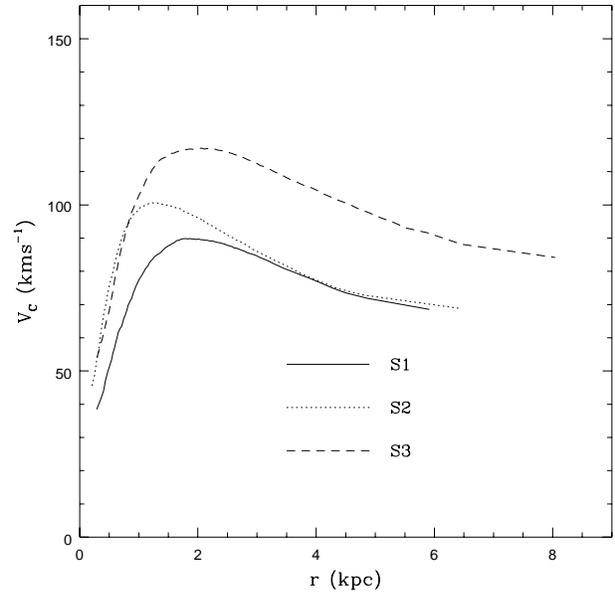}
\caption{ Rotation curve for our satellite models.}
\label{Fig. 2}
\end{figure}

\subsection{ Satellite Models}

The satellites are represented by self-consistent King models (King
1966) which provide a reasonable fit to early-type and nucleated dwarf
galaxies (Vader and Chaboyer 1994). These models are a sequence of
truncated isothermal spheres parametrized by a concentration $c \equiv
\log_{10}(r_t/r_c)$, where $r_t$ and $r_c$ are the so-called tidal and
core radii, respectively. To specify completely a satellite model we
provide the satellite mass, $M_S$, its concentration and its tidal radius. 
The latter is estimated from the density contrast
$\rho_S(r_t)/\overline{\rho}_G(r_a) \sim 3 $ at the apocentric
distance, $r_a$, of the orbit where the satellite is initially
placed. The parameters that characterize our satellite 
models, including the central density and the central one-dimensional 
velocity dispersion, are listed in Table 2 which are within the observed 
values (Vader and Chaboyer 1994; Binggeli and Cameron 1991; Bender, Burstein 
and Faber 1992). The rotation curve for each satellite model is shown in 
figure 2. We can notice that our satellite models are more concentrated than 
DD154-type dwarf galaxies (Moore 1994). In all cases, the 
satellite consists of $8192$ particles. \\

\subsection{ Numerical Methods and Orbital Parameters}

To follow the evolution of the galaxy-satellite system we use a tree
algorithm with a tolerance parameter of 
$\theta_{\hbox{{\footnotesize tol}}}=0.75$ and
an integration timestep of $1.3 \times 10^6$ yrs. Forces between particles are
computed including the quadrupole components (see Barnes and Hut (1986)
and Hernquist (1987) for details of this code). With these values, the
conservation of global energy and angular momentum is better than $1$ 
per cent for all our models.\\ 

Since our initial galaxy and satellite models are not in a perfect
equilibrium we allow them to relax separately before starting an
interaction simulation. The disc galaxy is relaxed for 20 time units, 
about one and a half rotation periods at its half-mass radius, 
while the satellite
is allowed to evolve for $40$ time units to reduce initial transient
effects. These two configurations are then superposed to create
our initial conditions. We have performed a set
of experiments varying the parameters that are likely to
influence the result of a galaxy-satellite interaction, for example, the
`circularity' of the orbit, the angle between the angular momentum of
the satellite and the disc, and the satellite structure. In Table 3
we list the parameters
of these simulations. Here, the `circularity' of the orbit has been
defined as $\epsilon_J \equiv  J/J_C(E)$, where $J$ is the
angular momentum of the satellite and $J_C(E)$ is the corresponding
angular momentum for a circular orbit of the same energy $E$ as the 
satellite's orbit. Also, we
have followed the subsequent evolution of the galaxy model in
isolation (our control model) to distinguish effects produced by
two-body encounters from those provoked by the accretion of the
satellite. The discussion of the following sections all refers to
satellites on orbits which actually intersect the disc. \\


\begin{table}
\begin{center}
\centering
\label {symbol3}
\caption{ Simulations.}
\begin{tabular}{|l|c|c|c|c|c|} \hline
 Name  &  Sat. model &  $\theta_i$ & $\epsilon_J$ & $r_p/R_D$ & $r_a/R_D$  \\ \hline \hline
G1S1    &    S1	     &  $45^{o}$  &   $0.33$   &  $1.5$ & $16.86$ \\
G1S2    &    S1	     &  $0^{o}$   &   $0.55$   &  $3$   & $15.71$ \\
G1S3    &    S1	     &  $45^{o}$  &   $0.55$   &  $3$   & $15.71$ \\
G1S4    &    S1	     &  $90^{o}$  &   $0.55$   &  $3$   & $15.71$ \\
G1S5    &    S1	     &  $135^{o}$ &   $0.55$   &  $3$   & $15.71$ \\ 
G1S6    &    S1	     &  $180^{o}$ &   $0.55$   &  $3$   & $15.71$ \\
G1S7    &    S1	     &  $0^{o}$   &   $0.82$   &  $6$   & $13.29$ \\
G1S8    &    S1	     &  $45^{o}$  &   $0.82$   &  $6$   & $13.29$ \\
G1S9    &    S2	     &  $0^{o}$   &   $0.55$   &  $3$   & $15.71$ \\
G1S10   &    S2	     &  $45^{o}$  &   $0.55$   &  $3$   & $15.71$ \\
G1S11   &    S2	     &  $90^{o}$  &   $0.55$   &  $3$   & $15.71$ \\
G1S12   &    S2	     &  $135^{o}$ &   $0.55$   &  $3$   & $15.71$ \\
G1S13   &    S2	     &  $180^{o}$ &   $0.55$   &  $3$   & $15.71$ \\
G1S14   &    S3	     &  $45^{o}$  &   $0.55$   &  $3$   & $15.71$ \\ 
G1S15   &    S3	     &  $135^{o}$ &   $0.55$   &  $3$   & $15.71$ \\
\hline
\end{tabular}
\end{center}
\medskip
$\theta_i$ refers to the angle between the initial angular momentum of
the satellite and the initial angular momentum of the disc. $\epsilon_J$ 
defines the circularity of the orbit, $r_p$ and $r_a$ correspond to
the initial pericentric and apocentric radii of the orbit, respectively. 
\end{table}



\begin{figure*}
\vskip 16cm
\caption{ The evolution of the disc for model G1S10. A fourth of the 
total number of disc particles has been plotted. The satellite has 
been tidally disrupted at time 252. By this time, the disc has been 
slightly tilted and shows an asymmetric configuration (see text for 
a discussion of some of these issues).}
\label{Fig. 3}
\end{figure*}


\begin{figure*}
\vskip 16cm
\caption{ The disc evolution for the retrograde counterpart of 
figure 3 (see model G1S12 in Table 3). The disc looks thicker 
than it really is because of tilting.}
\label{Fig. 4}
\end{figure*}


\begin{figure}
\plotone{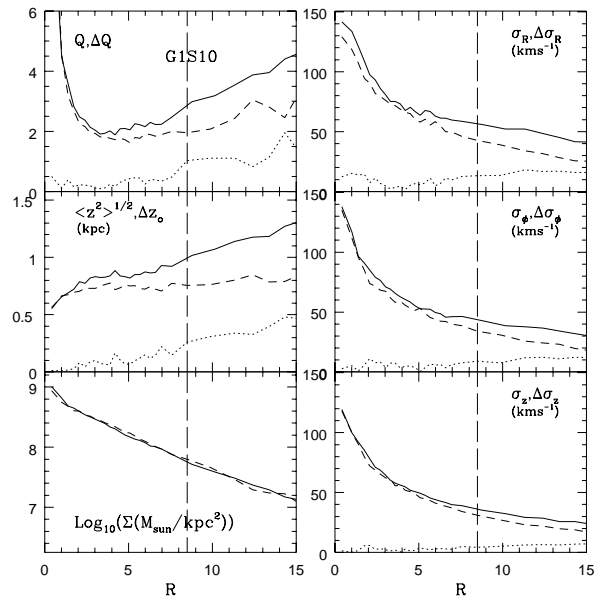}
\caption{ The kinematical properties of the disc in our
model G1S10. Solid lines give the kinematics of the disc in this 
model after $3.3$ Gyr, while short-dashed lines show corresponding 
quantities for our control model. Dotted lines show the 
difference between the two, and so the part of the changes due purely 
to the interaction. In all plots the vertical long-dashed line indicates 
the position at $R_\odot$.}
\label{Fig. 5}
\end{figure}

\section{ Disc Heating and Thickening }

In figures 3 and 4 we show the evolution of the disc of our models 
G1S10 and G1S12 (see Table 3). At the end of these simulations the 
satellite has been completely disrupted and the disc component has 
been altered in three different ways: (1) it is hotter and thicker; 
(2) it is no longer axisymmetric; and (3) it is tilted and warped. 
We discuss some of these effects in this and the following sections. \\


\begin{figure}
\plotone{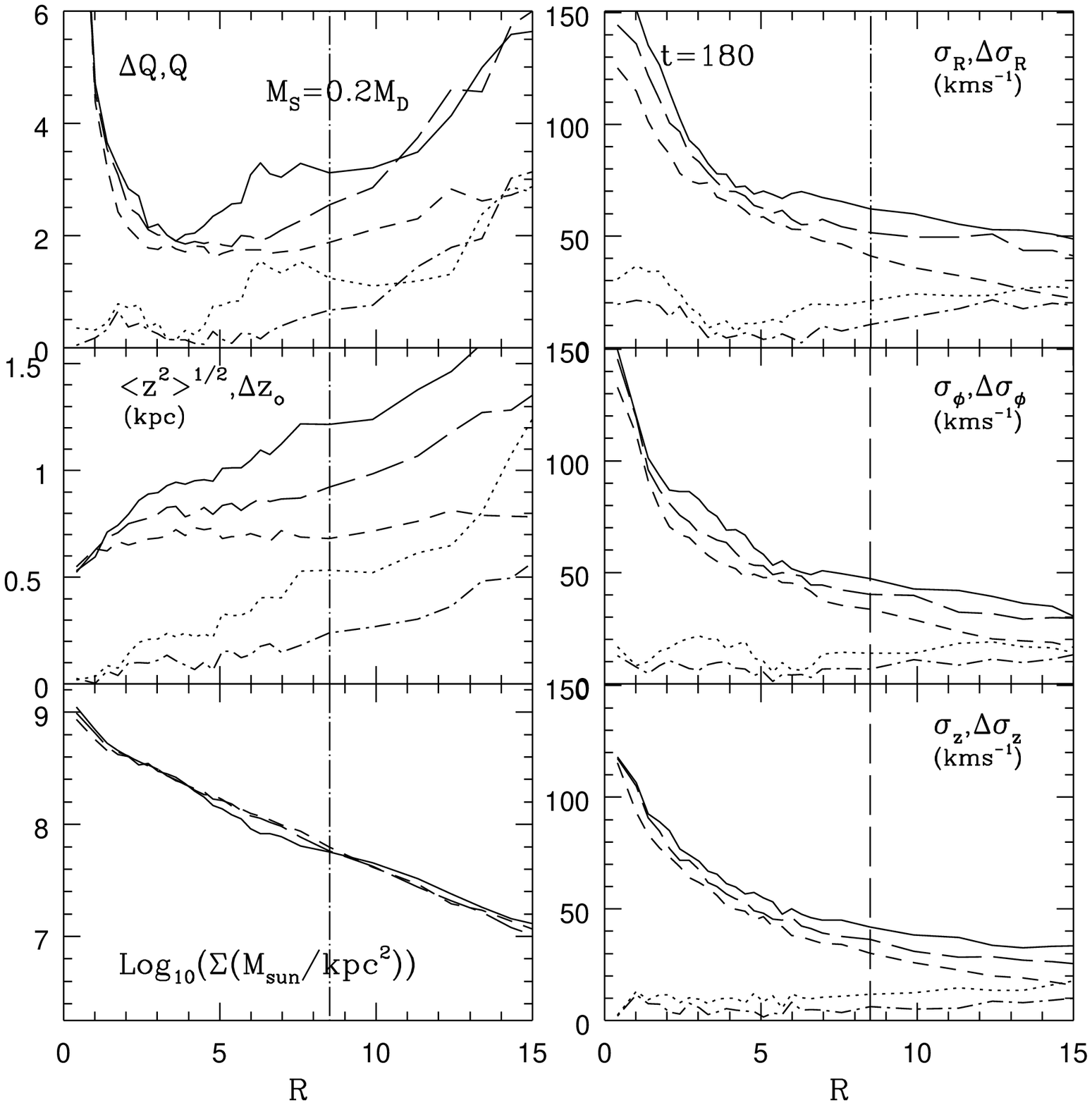}
\caption{ The response of the disc to the infall of
a satellite with an initial mass of $0.2\,M_D$. The satellite at this
time ($2.34$ Gyr) has been completely destroyed. The solid lines
correspond to a satellite following a prograde orbit with initial
angle of $\theta_i = 45^o$, (G1S14) while the long-dashed lines
correspond to its retrograde counterpart (G1S15) with $\theta_i =
135^o$. For reference, the
kinematical structure in our isolated control model is also
shown (short-dashed lines). The kinematical changes for model G1S14 
are indicated by dotted lines and for model G1S15 by dashed-dotted lines. 
The vertical line again indicates the position at the Solar radius.}
\label{Fig. 6}
\end{figure}

To determine the kinematics of the disc we rotate to axes aligned with
the principal
axes of the disc inertia tensor and then average particle properties in
concentric cylindrical annuli. These averages are a
crude measure of the disc kinematics since warping and
azimuthal asymmetry are not taken into account. The heating and thickening of the disc can be described by the
changes of the velocity dispersions $(\Delta \sigma_R, \Delta
\sigma_\phi, \Delta \sigma_z)$ and by the increase of the vertical
scale length, $\Delta z_o$. Effects due to two-body relaxation can
be subtracted by comparing the final state to that of the control model.
The increase of Toomre's stability parameter, $Q$, is also shown.
We compute the vertical scale length in each annulus 
using the following definition $z_o (R) \equiv <z^2>^{1/2}$. \\ 

In figure 5 we show the disc kinematical properties of our model G1S10 (solid lines)
together with the corresponding properties of the control model (short-dashed lines). It is evident 
from this figure that thickening of the stellar disc
does not occur uniformly at all radii; regions beyond $R_\odot$
are much more susceptible to damage by the infalling satellite than
the inner disc. Given the complexity of the
final disc structure we found it convenient to sample the kinematics at
three representative radii: near the centre (averaged out to $\sim 2$ kpc), at $R_\odot$, and at $14$ kpc. This is sufficient to provide us
with a global view of the heating and thickening. In Table 4 we give 
the final disc structure at these radii for all our simulations listed in 
Table 3\footnote{Satellite particles are not considered in computing the 
kinematical properties of the disc because, in most of the cases, their orbits 
do not fall into the disc plane and so they would be separeted from disc 
particles by their kinematics. Particles from satellites on prograde 
coplanar orbits would be more difficult to distinguish, however, in these 
simulations they form a disc-like structure with a hole of about $3-4$ kpc 
in the central region (see also Barnes 1996).}. For example, the velocity 
ellipsoid of the disc in model  
(G1S10) grows by $(18,12,8)$ kms$^{-1}$, $(12,10,5)$ kms$^{-1}$ and
$(19,14,8)$ kms$^{-1}$ (from inside to outside) resulting in 
increases of
Toomre's $Q$ parameter by about $0.4$, $0.8$ and, $2.2$. Before the accretion event $Q$ was roughly constant between $3$ kpc  
and $9$ kpc with a value of $\sim 1.5$ and after
satellite accretion it has risen to $Q \sim 2$ between $3$ kpc and $7$ kpc. 
The vertical scale length in this same case increases by $100$ pc, $275$ pc 
and $500$ pc, representing  a thickening by about $14$, $39$
and, $71$ per cent; while for its retrograde counterpart (model G1S11) is of $21$, $21$ and $43$ per cent. Disc thickening is even lower 
for model G1S3 being of $7$, $14$ and $43$ per cent (see entries in Table 4). 
In some of the models 
(G1S5, G1S10, G1S12 and G1S14) satellite accretion induces 
additional heating and thickening by exciting a bar-like instability 
in the inner $5$ kpc. \\ 

Three main trends are observed in our simulations: (1) a mighty
correlation between the change in disc structure and the relative
orientation of disc and satellite angular momenta. We find
that, in general, the disc is much more susceptible to damage by a satellite
on a prograde orbit than by its retrograde counterpart. This suggests
a resonant coupling between satellite orbit and disc. This difference
is most clearly seen by comparing the entries in Table 4 for 
the coplanar prograde case (model 
G1S9) and the corresponding polar case (model G1S11). In figure 6, we 
illustrate these effects using our two simulations with massive 
satellites (G1S14 and G1S15 in Table
3). Notice that for the retrograde case the disc is less strongly
affected at all radii. (2) In general, the planar
components of the velocity ellipsoid
respond more strongly than the vertical component to the
accretion of the satellite. Indeed, the ratio of the two planar components
is fixed by epicyclic theory. The degree of anisotropy 
is correlated with both the inclination and the sense of motion of the
orbit, being greatest for prograde coplanar orbits.
As a result Toomre's stability parameter $Q$ responds more
sensitively than the vertical scale length to the
accretion event. This is largely a consequence of the fact (addressed
in section 5) that
vertical perturbations tend to tilt the disc rather than to heat it.
(3) For our model G1S3 the effects of accretion on the disc are
quite modest, but a more massive and compact satellite can have a very
large effect (see entries of model G1S3 and G1S14 in Table 4). \\ 


\begin{table*}
\begin{center}
\centering
\caption{Disc kinematical changes.}
\begin{tabular}{|l|c|c|c|c|c|c|c|c|c|} \hline
\multicolumn{1}{|c|}{} & \multicolumn{3}{|c|}{$R_{c}^{*}$} & \multicolumn{3}{|c|}{$R_\odot$} & \multicolumn{3}{|c|}{$R_{4}^{\dag}$} \\ \hline
Model & $\scriptstyle{(\Delta \sigma_R,\Delta \sigma_\phi,\Delta \sigma_z)}$ & $\scriptstyle{\Delta Q}$ & $\scriptstyle{\Delta z_o}$ & $\scriptstyle{(\Delta \sigma_R,\Delta \sigma_\phi,\Delta \sigma_z)}$ & $\scriptstyle{\Delta Q}$ & $\scriptstyle{\Delta z_o}$ & $\scriptstyle{(\Delta \sigma_R,\Delta \sigma_\phi,\Delta \sigma_z)}$ & $\scriptstyle{\Delta Q}$ & $\scriptstyle{\Delta z_o}$ \\ 
	  &    (kms$^{-1}$) &        &    (pc)& (kms$^{-1}$)&        &  (pc)
& (kms$^{-1}$) &       &  (pc) \\ \hline \hline
{G1S1}    &    $(9,6,4)$  & $0.3$  &  {$50$}  & $(5,3,2)$  & $0.5$   &  {$150$}  & $(10,7,4)$   & $1.2$ & $225$ 	\\
{G1S2}    &    $(16,8,4)$ & $0.4$   &  {$75$}  & $(16,10,4)$ & $1.1$  &  {$200$}  & $(32,21,6)$ & $2.6$ & $300$ 	\\
&    [$(16,7,5)$] & [$0.5$]   &  [{$100$}]  & [$(31,12,4)$] & [$1.4$]  &  [{$200$}]  & [$(47,26,9)$] & [$2.9$] & [$450$]     \\
{G1S3}    &    $(10,7,5)$  & $0.3$   &  {$50$}  & $(5,3,2)$  & $0.5$   &  {$100$}  & $(11,6,5)$ & $1.4$  & $300$    	\\
{G1S4}    &    $(12,6,4)$  & $0.5$  &  {$100$}  & $(4,3,2)$   & $0.4$  &  {$100$}  & $(6,5,4)$  & $0.9$ & $250$  	\\
{G1S5$^{\ddag}$}    &    $(24,16,8)$   & $0.6$  &  {$200$}  & $(4,2,3)$  & $0.5$   &  {$100$}  & $(6,6,4)$  & $1.0$ & $150$  	\\ 
{G1S6}    &    $(6,5,7)$  & $0.3$ &  {$75$}  & $(5,4,3)$  & $0.4$ &  {$100$}  & $(10,7,5)$ & $1.7$ & $275$  	\\
{G1S7}    &    $(38,18,8)$  & $0.4$ &  {$100$}  & $(16,12,5)$  & $1.2$ &  {$250$}  & $(27,14,7)$ & $3.4$ & $450$  	\\
    &    [$(38,17,9)$]  & [$0.4$] &  [{$100$}]  & [$(15,12,6)$]  & [$1.1$] &  [{$250$}]  & [$(29,16,8)$] & [$3.4$] & [$500$] \\
{G1S8}    &    $(10,10,6)$  & $0.3$  &  {$50$}  & $(10,7,4)$  & $0.8$  &  {$150$}  & $(17,12,6)$ & $2.2$ & $350$  	\\
{G1S9}    &    $(19,13,11)$  & $0.5$ &  {$100$}  & $(22,15,7)$  & $1.5$ &  {$300$}  & $(37,23,9)$  & $2.6$ & $550$  	\\
       &   [$(19,13,11)$]  & [$0.5$] &  [{$100$}]  & [$(25,15,8)$]  & [$1.6$] &  [{$350$}]  & [$(43,28,12)$]  & [$2.9$] & [$675$]  	\\
{G1S10$^{\ddag}$}    &    $(18,12,8)$  & $0.4$ &  {$100$}  & $(12,10,5)$  & $0.8$ &  {$275$}  & $(19,14,8)$  & $2.2$ & $500$  	\\
{G1S11}   &    $(11,10,6)$  & $0.4$ &  {$100$}  & $(7,4,4)$ & $0.5$ &  {$150$}  & $(11,8,5)$ & $1.0$ & $300$  	\\
{G1S12$^{\ddag}$}    &    $(17,11,11)$  & $0.8$ &  {$150$}  & $(4,4,4)$ & $0.4$ &  {$150$}  & $(10,6,6)$ & $1.0$ & $300$  	\\ 
{G1S13}    &    $(7,8,4)$  & $0.5$ &  {$75$}  & $(12,10,7)$ & $0.8$ &  {$250$}  & $(19,15,10)$ & $2.4$ & $700$  	\\
{G1S14$^{\ddag}$}    &    $(32,20,12)$  & $0.6$ &  {$150$}  & $(22,15,12)$ & $1.2$ &  {$550$}  & $(27,18,16)$ & $2.9$ & $800$  	\\
{G1S15}    &    $(20,12,8)$  & $0.6$ &  {$100$}  & $(11,9,6)$ & $0.8$ &  {$300$}  & $(21,12,10)$ & $2.2$ & $525$  	\\ \hline

\end{tabular}
\end{center}
\medskip
Symbols $*$ and $\dag$ denote quantities at the centre and $4\,R_D$,
respectively. Models forming a bar are indicated by the symbol
$\ddag$. Quantities in brackets are computed by taking into account satellite 
particles.
\end{table*}

\subsection{ Disc Heating and the Bulge/Disc Mass Ratio}

To address the effect of the bulge in the heating process of the stellar 
disc we have repeated a few of our simulations but with two additional 
galactic models. One of them has a bulge with a mass of $M_B=0.2\,M_D$ and the other with a mass of $M_B=2/3\,M_D$. We did not pursue a larger reduction of 
the bulge mass since it requires a huge number of particles $(> 5\times 10^5)$ to keep stable the disc against bar formation during the 
disruption times in our simulations; this would be prohibitive (e.g. Walker et al. 1996). In these models both the disc and halo are 
the same as in our primary galaxy model of section 2. In figure 1 we 
show the rotation curves for these new galactic models.\\

In Table 5 we summarize the disc kinematical changes for these new experiments. 
Prefix G2 and G3 refer to the galactic model with the less and more 
massive bulge, respectively. We should mention that model `G2' develops a 
bar in isolation at time $\sim 3.2$ Gyr inside $4$ kpc. Comparing the entries 
in tables 4 and 5 at the Solar radius we can see that the disc has 
been thickened by $39$, $39$ and $21$ per cent for models G1S10, G2S10 
and G3S10, respectively, while for their retrograde counterparts (G1S12, G2S12 and G3S12) the corresponding values are only of $21$, $29$ and 
$14$ per cent. This suggest that a very 
massive bulge would be more efficient in reducing the heating and 
thickening of the stellar disc. This effect can be more clearly appreciated in 
our models with the more massive satellite (models G1S14, G1S15, G3S14 
and G3S15). Thus, the difference between the thickening found 
by Walker et al. (1996) ($\sim 60$\%) and that in our simulations can 
be attributed 
to the fact that in the former case the satellite interacts strongly with 
the disc because of the orbit chosen and because of the bulgeless galaxy 
model. The latter forms a bar during the early stages of the accretion. \\   


\begin{table*}
\begin{center}
\centering
\caption{Disc kinematical changes.}
\begin{tabular}{|l|c|c|c|c|c|c|c|c|c|} \hline
\multicolumn{1}{|c|}{} & \multicolumn{3}{|c|}{$R_{c}^{*}$} & \multicolumn{3}{|c|}{$R_\odot$} & \multicolumn{3}{|c|}{$R_{4}^{\dag}$} \\ \hline
Model & $\scriptstyle{(\Delta \sigma_R,\Delta \sigma_\phi,\Delta \sigma_z)}$ & $\scriptstyle{\Delta Q}$ & $\scriptstyle{\Delta z_o}$ & $\scriptstyle{(\Delta \sigma_R,\Delta \sigma_\phi,\Delta \sigma_z)}$ & $\scriptstyle{\Delta Q}$ & $\scriptstyle{\Delta z_o}$ & $\scriptstyle{(\Delta \sigma_R,\Delta \sigma_\phi,\Delta \sigma_z)}$ & $\scriptstyle{\Delta Q}$ & $\scriptstyle{\Delta z_o}$ \\ 
	  &    (kms$^{-1}$) &        &    (pc)& (kms$^{-1}$)&        &  (pc)
& (kms$^{-1}$) &       &  (pc) \\ \hline \hline
{G2S10$^{\ddag}$}    &    $(16,11,10)$  & $0.7$  &  {$75$}  & $(12,7,7)$  & $0.8$   &  {$275$}  & $(18,11,9)$   & $1.3$ & $600$ 	\\
{G2S12$^{\ddag}$}    &    $(9,10,5)$ & $0.3$   &  {$100$}  & $(7,5,4)$ & $0.7$  &  {$200$}  & $(11,6,9)$ & $0.7$ & $400$ 	\\ \hline
{G3S3}    &    $(5,6,6)$  & $0.6$   &  {$50$}  & $(4,3,3)$  & $0.4$   &  {$100$}  & $(7,6,4)$ & $0.8$  & $200$    	\\
{G3S10}    &    $(9,6,5)$   & $0.8$  &  {$100$}  & $(7,5,5)$  & $0.5$   &  {$150$}  & $(16,10,7)$  & $1.5$ & $400$  	\\ 
{G3S12}    &    $(6,5,3)$  & $0.7$ &  {$50$}  & $(4,3,2)$  & $0.4$ &  {$100$}  & $(6,6,5)$ & $0.7$ & $275$  	\\
{G3S14}    &    $(16,6,9)$  & $0.8$  &  {$75$}  & $(10,8,8)$  & $0.8$  &  {$325$}  & $(26,14,13)$ & $2.5$ & $750$  	\\
{G3S15}    &    $(10,8,4)$  & $0.5$  &  {$50$}  & $(6,4,5)$  & $0.5$  &  {$200$}  & $(13,9,8)$ & $1.4$ & $425$  	\\ \hline
\end{tabular}
\end{center}
\medskip
Symbols as in Table 4.
\end{table*}


\begin{figure*}
\ppplotone{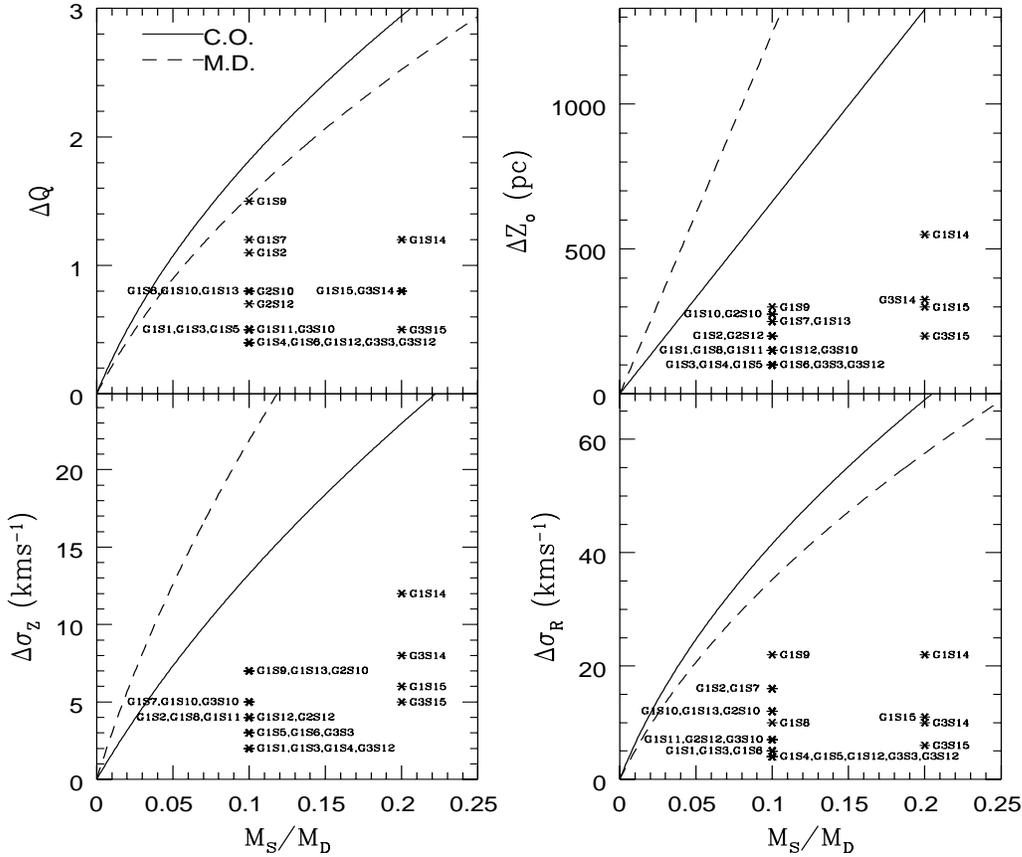}
\caption {Disc kinematical changes in the Solar neighborhood. Here, we
show the changes in the parameter $Q$ (top-left), in the vertical
scale length (top-right), and in the vertical (bottom-left) and radial
(bottom-right) velocity dispersions as resulting by the accretion of a
satellite of mass $M_S$. Solid and dashed lines represent the
kinematical changes as predicted by TO's formulae while starred 
symbols represent the results from all our simulations (see text for details).}
\label{Fig. 7}
\end{figure*}

\section{ Comparison with TO's Results}

It is interesting to compare the size of the effects we find with
those predicted analytically by TO. They give formulae for the changes
in the disc vertical scale length, velocity dispersions and stability
parameter $Q$ (their equations 2.16, 2.21, 2.25 and 2.26). These can be
used to estimate the quantities we list in Table 4 and 5. We apply
these formulae by inserting the properties of the unperturbed disc (as
measured in our control model) at the Solar radius and the satellite
total mass, and by averaging over the orbital inclination of the
satellite. In figure 7,
we show the resulting disc changes as predicted by TO
(solid and dashed lines for two extreme cases) together with 
those found in our simulations (denoted by starred symbols). The solid 
lines correspond to a vertical ($\epsilon_\perp$) and 
epicyclic ($\epsilon_\parallel$) efficiency heating assuming circular 
orbits with random inclinations for the 
velocity distribution of the satellites while the dashed lines refer to 
an isotropic Maxwellian distribution (see TO, Appendix A). Clearly, the 
analytical predictions tend to
overestimate the disruptive effects of satellite accretion by a
non-negligible factor; typically about $2-3$. These
differences are more marked for the more massive satellite. Notice
that for models G1S2, G1S7 and G1S9 the change in
the stability parameter is closer to that predicted by TO. This is
because the satellite in these models (which follows a prograde orbit
near the disc plane) induces substantial deviations in $\Sigma_D(R)$. These
deviations were taken into account in determining the change in $Q$
for figure 7. If we ignore such variations,
as TO did, then $\Delta Q$ is entirely determined by $\Delta
\sigma_R$; the difference is then more dramatic since
the change in $Q$ is reduced to $0.8$, $0.7$ and $1.0$ for models G1S2, 
G1S7 and G1S9, respectively. \\

The fact that we find substantially weaker effects than those
predicted by TO suggests that their limits on the accretion rate of
satellites (based on the thinness of observed discs, and their
susceptibility to spiral instabilities) are likely to be too strict.
There are some additional effects which may result in our own results
still being a significant overestimate of the extent of the damage.
\begin{itemize}
\item The stellar dynamical heating of the disc in our 
simulations may be partially compensated by dissipative effects in a
gaseous component. N-body/Hydrodynamical simulations carried out by Mihos and
Hernquist (1994) and the recent work by HC reveal that gas cooling can make
an important contribution to sustaining spiral structure in galaxies.
\item Our simulations begin with the satellite already
quite close to the disc and have relatively low mass halos. They are
thus likely to underestimate disruptive effects on the satellite
before it begins to interact strongly with the disc (see HC).
Progress on this point will require larger simulations and properly
realistic ``cosmological'' initial conditions. Currently available 
results suggest that discs may accrete material less efficiently
than haloes (NFW94, NFW95). 
\item Damaging effects on disc structure will be
diminished whenever the satellite is less concentrated than those we
have considered and so can be disrupted before encountering the disc
(again see HC).  Although simulations from
cosmological initial conditions show that the distribution of orbital 
eccentricities should be almost uniform so that a significant
fraction of satellites reach the disc essentially unperturbed on their first
passage, such
high velocity, near-radial encounters cause less damage than
prograde encounters at lower velocities.
\end{itemize}
It is interesting that the properties of the most strongly perturbed
discs in our simulations do seem to agree with those of the thick disc
component at the solar radius. This component has a velocity ellipsoid
of about $(63,42,38)$ kms$^{-1}$, a vertical scale length of $1$ kpc
and an asymmetric drift of about $25$ kms$^{-1}$ (Freeman
1996). Similar structure can be seen in the
galaxy NGC $4565$ for which the vertical scale length is $\approx \,
1.5$ kpc (Morrison 1996). \\

The difference between TO's analytical results and those of our
simulations may be due to one or a combination of the following
factors: 
\begin{itemize}
\item A self-gravitating satellite has the capacity to absorb part of
its orbital energy which can be carried away by the stripped
material. In this way, the deposition of energy onto the disc is diminished. 
\item Another important ingredient in the galaxy-satellite interaction
is the responsiveness of the halo. This regulates the transfer of
angular momentum between the different components and may weaken the
resonant response of the disc to the infalling satellite.
\item In addition, TO do not make any distinction between satellites
on prograde and retrograde orbits and this has important consequences 
for the survival and evolution of the disc (see section 3 and 5).
\item Another controversial point in TO's derivation is their
assumption that the deposition of a satellite's orbital energy in the
disc is a local process. They argued that the overall energy change in
the disc would be the same if the excitation of modes
in the disc were taken into account, but the formation of spirals and
warps reveals that {\sl global} modes are excited by
the infalling satellite before it crosses the main disc.
\end{itemize}


\begin{figure}
\plotone{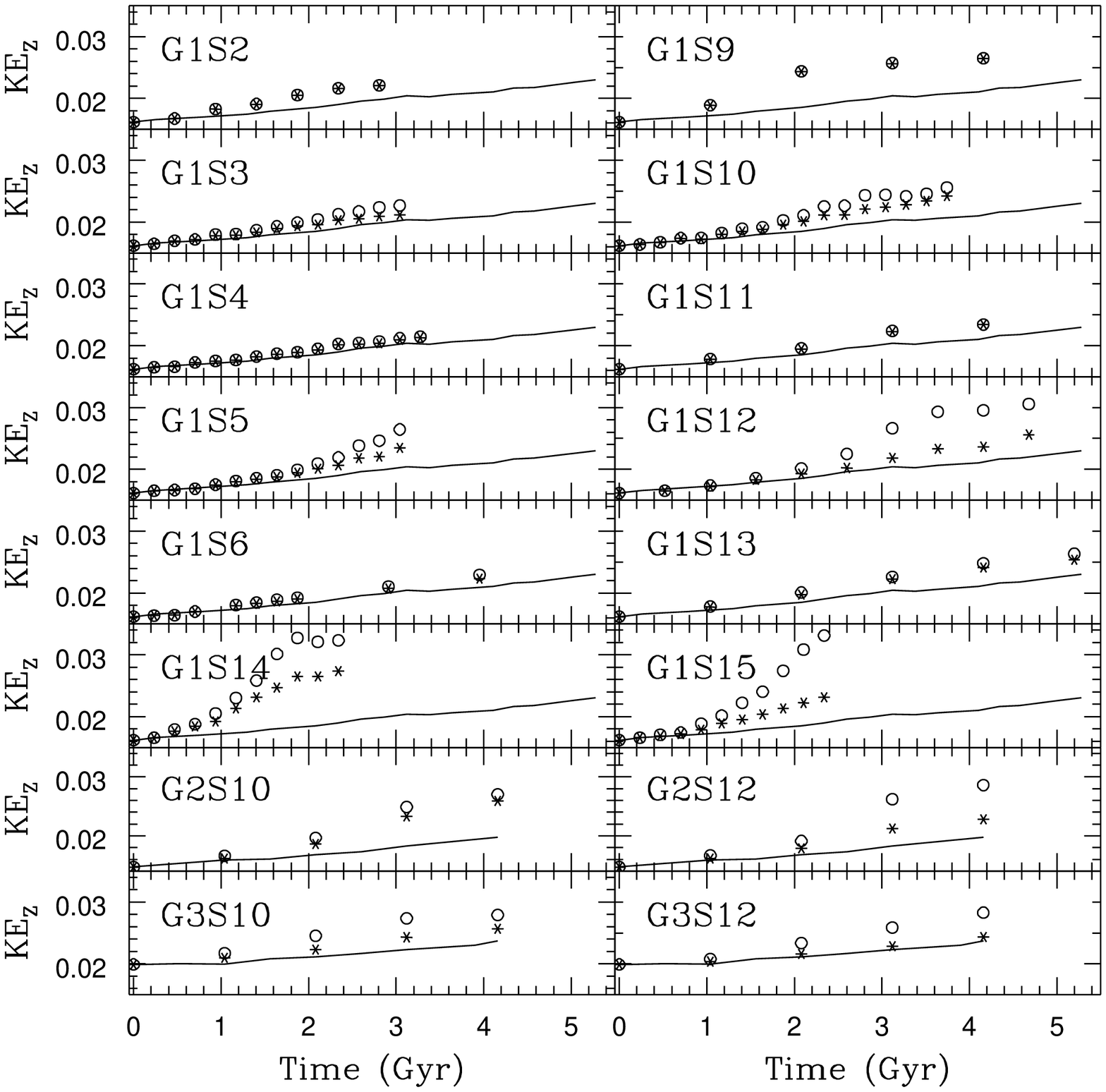}
\caption{Evolution of the vertical kinetic energy of the
disc for several of our models.
The solid line represents the vertical heating
produced by numerical relaxation. The starred symbols refer to the
vertical kinetic energy as measured from a frame that coincides with
the main axes of the inertia tensor while the open circles show its
evolution in the initial reference frame.} 
\label{Fig. 8}
\end{figure}

\section{ Disc Tilting and Warping}

As the satellite sinks to the galactic centre it transfers energy and
angular momentum to its surroundings (QHF, HC). Part of this energy,
as we saw in section 3, goes to the disc in random vertical motions
giving rise to the disc thickening. There is also, however, a coherent
response of the disc to the satellite accretion which is associated
with the tilting observed in figures 3 and 4. Looking in figure 8 we can also
appreciate the effect of the orbital type, compactness and mass of the 
satellite and the mass of the bulge on the evolution of the vertical 
kinetic energy. The solid line corresponds to the
vertical heating of the disc component coming from numerical
relaxation while the starred symbols refer to the vertical kinetic
energy as measured in a reference frame that coincides with the main
axes of the inertia tensor and the open circles correspond to the
vertical kinetic energy in the original reference frame. We can appreciate 
some general trends: (i) In the prograde models most of the energy 
goes to thickening 
the disc which is best illustrated by models G1S2, G1S9 and G1S14. 
In contrast, the 
retrograde models clearly show a coherent response of the disc (e.g. G1S12 and 
G1S15). Thus, a smaller fraction of the energy goes to thickening the disc in 
this case which tell us that the disc is more robust to mergers with satellites 
on retrograde orbits than on prograde ones. Notice that for satellites on 
polar and retrograde coplanar orbits all the energy goes to the disc, 
however, the 
thickening is lower than in their prograde counterparts suggesting that the 
vertical kinetic energy is distributed in the satellite remnants and 
halo particles. (ii) As pointed out in section 3, a more massive bulge 
may reduce the vertical heating of the disc which is more dramatic in models 
G1S10, G2S10 and G3S10 but this is not the case in their 
retrograde counterparts for which the coherent response of the disc seems to 
be reduced (e.g. models G1S12, G2S12 and G3S12). \\


\begin{figure*}
\vskip 18cm
\caption{Representation of the disc by a set of rings equally spaced
in the polar radius. Each ring has been displaced to the centre
defined by the centre of mass of the particles contained in it and
then rotated to the principal axes of its inertia tensor. The size of
the boxes are $56 \times 56 \times 3.5$ in kpc so the plotted distortions 
from planarity exaggerate the true distortions. The mean radii of the
rings shown in this figure are: $4.37$, $7.88$, $11.38$, $14.87$, and
$18.35$ kpc.}
\label{Fig. 9}
\end{figure*}

In the absence of a satellite there is a small exchange of angular
momentum between the disc and the halo resulting in a small tilt of the
disc, less than $3^o$ over $\sim 4.5$ Gyr. When a satellite is
introduced in the galaxy, the transfer of angular momentum changes
dramatically. By the end of the simulation, the disc of model G1S14 
has been tilted by $\sim 11^o$ while the angle $\theta_i$ (for both bound
and unbound satellite particles) has fallen by $\sim 26^o$ from 
its initial value. For model G1S15, the disc has been tilted by $\sim
15^o$, which, as noted above, is larger than for the corresponding
prograde case. In this last model, the angle $\theta_i$ remained
almost unchanged which tell us about the importance of the
coupling between disc stars and the satellite orbit. However, in both
cases about $45$ per cent of the total initial angular momentum of the
satellite remains in the satellite debris, the rest going to the
disc and the halo. Less massive satellites experience less dramatic
evolution since the sinking rate and, hence, 
the loss of angular momentum is proportional to the square of the
satellite mass. Thus, in our models G1S3 and G1S5 about 61\% and 68\%,
respectively, of the angular momentum remains in the satellite 
debris. Also the disc tilting is only about $6^o$ and $8^o$ in these
cases. \\

On the other hand, observations of edge-on disc galaxies show that a
considerable fraction of them have a conspicuous feature in the form of
an `integral sign', commonly known as a warp. Most of these structures
develop far beyond the optical disc and are very extended (Briggs
1990). A typical example is our own Galaxy for which the HI warp,
detected at $21\,$cm, is revealed beyond the solar circle  (Henderson
et al. 1982). However, HI warps are observable mainly where the
optical disc ends and the detection of stellar warps is harder. A
systemic attempt was made by S\'anchez-Saavedra, Battaner and Florido
(1990) by selecting a sample of $82$ edge-on disc galaxies from the
Palomar Survey. They were able to identify $23$ cases of stellar warps
in that sample. This study suggests that stellar warps are also a
common feature, but their origin remains a puzzle.\\ 

From a theoretical perspective, several scenarios have been proposed to
try to explain the origin of the warping of disc galaxies (an
excellent review on the subject is found in Binney 1992). Among the
most frequently invoked is a cosmological origin reflecting the time when
the galaxy was built up. Current cosmological simulations suggest that
dark haloes are highly flattened with mean axis ratios of $<c/a> \sim
0.5$ and $<b/a> \sim 0.7$ (Dubinski and Carlberg 1991). As a
consequence, the disc angular momentum may not be in
alignment with one of the main halo axes which may then induce 
a warping mode in the disc. This tilting mode was studied
by Sparke and Casertano (1988) for the case of a disc embedded in a
flattened rigid halo; they found that it is consistent with observed 
warps. However, recent self-consistent N-body simulations
carried out by Dubinski and Kuijken (1995) have shown that the
response of the halo can not be ignored, since dynamical
friction can damp the warps in much less than a Hubble time. Nelson and
Tremaine (1995) used perturbation theory to arrive at a similar
conclusion, suggesting that, in general, a primordial origin of
galactic warps may be ruled out.\\ 

Another mechanism to excite and maintain warps resorts to the accretion
of satellites (Binney 1992) which has been highlighted in a recent
paper by Weinberg (1996). This last author has shown the importance
of the response of the halo (assumed spherical) of our Galaxy to the
interaction with the LMC, and its subsequent effect on the formation of
the disc's HI warp. Since in all our simulations the halo is assumed 
spherical we can address the relevance of such events in the formation
of stellar warps. For this, we proceeded by rotating our disc along
the main axes of the inertia tensor of the particles located inside
$3\,R_D$ and representing it by a system of initially concentric
rings equally spaced and each containing a sufficient number of
particles. This done, each ring is displaced to a new centre
determined by the centre of mass of the particles within it. Finally,
we compute the inertia tensor for each ring and \mbox{rotate} to its principal 
axes. Figure 9 shows the result of this process for our isolated
galaxy model `G1' (upper panel) and for our models G1S3, G1S8 and G1S14,
respectively. The size of the each box is $56 \times 56 \times 3.5$ in
kpc. We can see that model G1S3 produces no 
distinguishable warp while a
more massive satellite (model G1S14) in the same orbit or a 
satellite on
a more nearly circular orbit (model G1S8)
produces bigger changes. However, it is important to remark that even
in the most favorable case the departure of the warp from the disc
plane is less than $1.75$ kpc in the outer parts which 
correspond to an angle less than $7^o$.\\

\section{ Satellite Sinking and Disruption Times}

In order to verify the reliability of the decay rate and disruption
time of the satellites in our simulations we built up a self-consistent
model called 'NEW' similar to our model G1S3 but with only a fourth
of the number of particles in each of the components present in the
model. The mass evolution and
disruption rate of the satellite for these models are shown in figure
10(a)-(b) where the open circles represent model G1S3 and
the open squares joined by solid lines correspond to the 'NEW'
model. Notice that the agreement is good with differences in the
disruption process of the satellite being more evident in the last
stages of its evolution.\\ 


\begin{figure}
\plotone{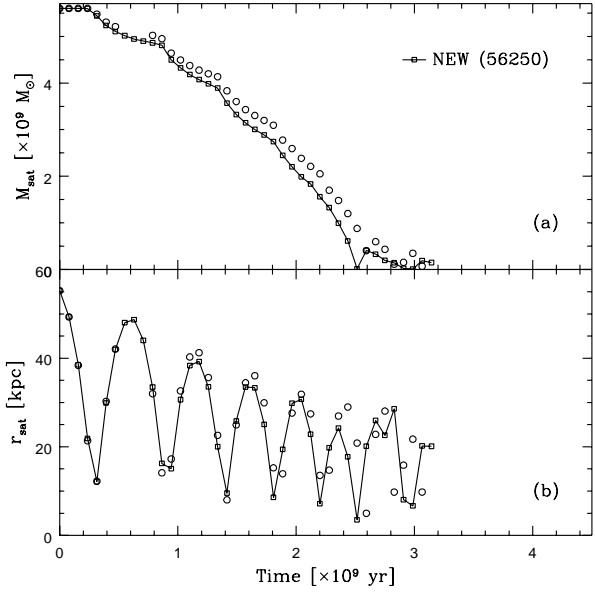}
\caption{(a) This figure shows the mass of the satellite as a function
of time. In both models G1S3 (open circles) and NEW we obtained similar 
times for the total disruption of the satellite which occurs about $2.6-2.8$
Gyr. (b) The satellite position is presented as a function of
time. Notice that the satellite is completely destroyed before it
reachs the centre of the galaxy. }
\label{Fig. 10}
\end{figure}


\begin{figure}
\plotone{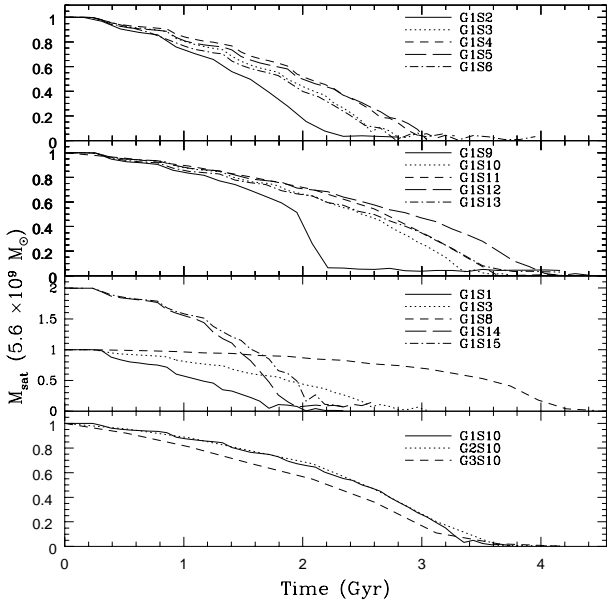}
\caption{The disruption times of the satellites for most of our
simulations. Most of the damage to the satellites occurs at pericentre 
since the tidal force due to the galactic potential is strongest there.}
\label{Fig. 11}
\end{figure}


\begin{figure}
\plotone{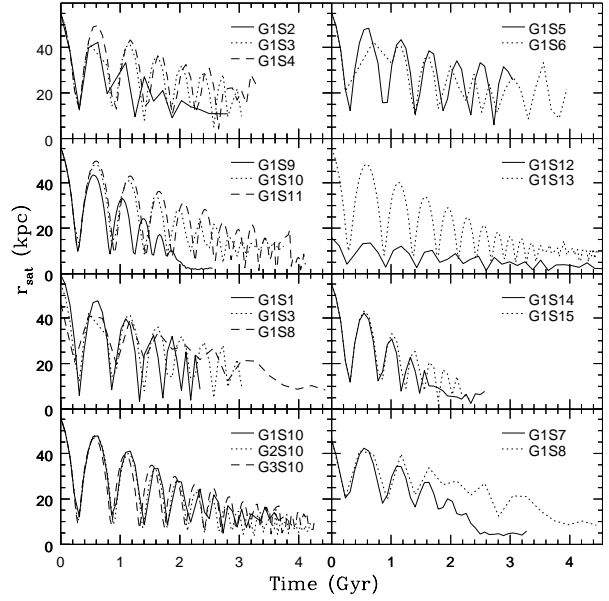}
\caption{The orbital decay rate for the satellites 
in most of our simulations.}
\label{Fig. 12}
\end{figure}

As the satellite is being accreted by the host galaxy, there are two main 
physical \mbox{mechanisms} that regulate its evolution: the tidal
interaction and dynamical friction. Figures 11 and 12 show the effects
of these mechanisms in the overall evolution of the satellites for most of 
our simulations. These  effects are illustrated by
our model G1S1. Notice that the biggest `jumps' in the mass evolution
of the satellite correspond to its passage through pericentre. This is
easily understood since at that position the tidal interaction of the
satellite with the host galaxy is the strongest since the
gradient of the galactic gravitational force reachs a maximimum at
pericentre. Comparison of the different models yields the following
conclusions: 
(i) for a given circularity and satellite model, the disruption of the
satellite occurs faster for prograde orbits than for retrograde
ones being more evident for coplanar orbits. Furthermore, just from 
the mass loss and disruption total times, it is
difficult to distinguish between polar and retrograde orbits. In the
case of prograde orbits, the satellites suffer a faster disruption and
sinking if they follow orbits which are closer to coplanar. 
(ii) For a given orbit, the decay rate and disruption of our satellite 
with an initial mass of $0.2\,M_D$ are faster than for the satellite with an 
initial mass of $0.1\,M_D$ since the decay rate is proportional to the
satellite mass. (iii) For a given satellite mass and orbit a
more compact satellite survives for a longer time than a less
compact one since its binding energy is greater. (iv) Finally,
although we do not have a large number of simulations, we note that, 
for a given satellite, the disruption time scales roughly as 
$\epsilon_J^{1/3}$.\\

\section{ Chandrasekhar's Dynamical Friction Formula}

In this section we briefly address the reliability of Chandrasekhar's
dynamical friction formula (Chandrasekhar 1960) for describing the decay
rate of the satellites in our N-body simulations. To carry
out this study, we built up a galaxy model similar to our fully
self-consistent one but replacing the self-gravitating halo by a rigid
potential which is free to move in response to the particle
distribution of the other components. For a Maxwellian distribution of
velocities with a dispersion velocity $\sigma(r)$ we have that the
dynamical friction on the satellite is given by (Binney and Tremaine
1987): 

\begin{equation}
{\bf F}_{df}\,=\,-{{4\pi \ln \Lambda G^2 M_S^2 \rho_H(r)}\over{v_S^3}}[\hbox{ erf}(X)-{{2 X}\over{\sqrt{\pi}}}\hbox{ e}^{-X^2}]\,{\bf v}_S,
\end{equation} 

\noindent where $X \equiv {v}_S/(\sqrt{2}\sigma(r))$ and $\ln \Lambda$ is the
Coulomb logarithm. The basic underlying assumption in the derivation
of this equation 
is that a point particle of mass $M_S$ moves with a velocity ${\bf
v}_S$ in an homogeneous and infinite background of lighter
particles whose self-gravity is completely ignored. As this point
particle travels across the background it deflects particles
(gravitational focusing) producing a density enhancement (a wake)
behind it which is responsible for the drag force expressed in equation
(6) (Mulder 1983). Obviously, the manifestation of this force requires
a halo made of independent particles and hence, it will be not present in a
rigid model. Since we want to check the reliability of equation (6) to
describe the sinking rate in our self-consistent simulations we have
introduced it `by hand' in the case of the rigid halo models to
emulate a drag force acting on the satellite. This will allow us to
determine the importance of the self-gravity of the background particles of 
the halo particles to the sinking process. We should remark that the
disc was kept alive and, hence, `disc friction' is included. \\ 


\begin{figure}
\plotone{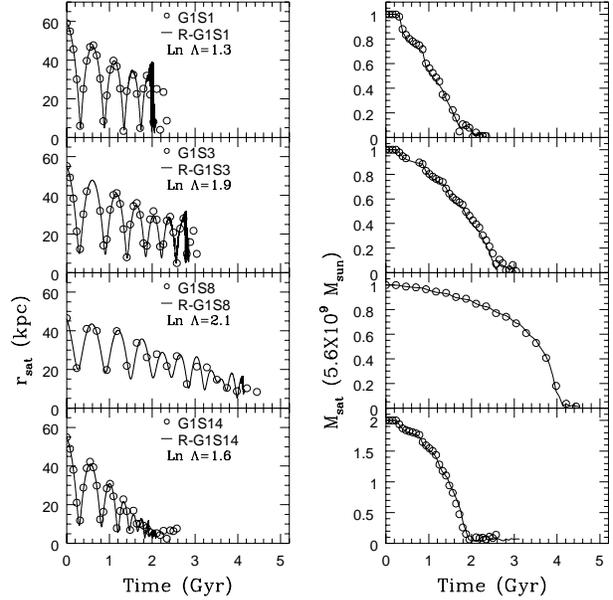}
\caption{Decay and disruption rates for a few of our models. Open
circles corresponds to our fully self-consistent simulations while
solid lines represent the same model but with a rigid halo free to
move instead. Dynamical friction has been computed using
Chandrasekhar's relation.} 
\label{Fig. 13}
\end{figure}

In addition to the study of the orbital decay of satellites within
their host galaxies (e.g. QG, QHF) dynamical friction has been invoked to 
explain the evolution of cD galaxies in clusters 
of galaxies (Ostriker and Tremaine 1975, White 1976) as well as 
the formation of galactic nuclei (Tremaine, Ostriker and
Spitzer 1975). Despite numerous studies, there is
still disagreement about the applicability of Chandrasekhar's relation in 
such situations and the relevance of its inherent local character. 
Thus, Lin and Tremaine (1983) using a semi-restricted N-body code found
that equation (6) provides accurate decay rates for satellites
following circular orbits for a large part of the parameter space
(satellite mass, satellite size, and number density and mass of halo
stars) suggesting that the self-gravity of the halo stars is
unimportant. In contrast, White (1983) using fully self-consistent N-body
calculations based on a spherical harmonics expansion found that a
local description of the dynamical friction was not enough to
determine the sinking times in satellites. However, Bontekoe and van
Albada (1987) reached a conclusion more in agreement with Lin and
Tremaine (1987) since they found that higher order terms of the
harmonic expansion had no discernible effect on their results. In an
attempt to clarify this situation, Zaritsky and White (1988) carried
out an exhaustive study using several codes and found that the
self-gravity of the halo is not important for the sinking rate of the
satellite. Also, Hernquist and Weinberg (1989) used self-consistent
and semi-restricted N-body algorithms to study this problem and to try
to understand the discrepancy. They concluded that orbital decay is
strongly suppressed if self-gravity of the response is taken into
account in disagreement with Zaritsky and White (1988). Since most of
the previous work was restricted to satellites following circular
orbits, S\'eguin and Dupraz (1994, 1996) employed a semi-restricted
method to determine the applicability of Chandrasekhar's relation to
head-on encounters. They also conclude that the global response of the
galaxy cannot be ignored. \\ 

\begin{table*}
\begin{center}
\centering
\caption{Disc kinematical changes. Rigid haloes}
\begin{tabular}{|l|c|c|c|c|c|c|c|c|c|} \hline
& \multicolumn{3}{|c|}{$R_{c}^{*}$} & \multicolumn{3}{|c|}{$R_\bullet$} & \multicolumn{3}{|c|}{$R_{4}^{\dag}$} \\ \hline
{Model} & $\scriptstyle{(\Delta \sigma_R,\Delta \sigma_\phi,\Delta \sigma_z)}$ & $\scriptstyle{\Delta Q}$ & $\scriptstyle{\Delta z_o}$ & $\scriptstyle{(\Delta \sigma_R,\Delta \sigma_\phi,\Delta \sigma_z)}$ & $\scriptstyle{\Delta Q}$ & $\scriptstyle{\Delta z_o}$ & $\scriptstyle{(\Delta \sigma_R,\Delta \sigma_\phi,\Delta \sigma_z)}$ & $\scriptstyle{\Delta Q}$ & $\scriptstyle{\Delta z_o}$ \\ \hline \hline
R-G1S1    &    $(9,6,4)$  & $0.4$  &  $50$  & $(5,4,5)$  & $0.4$   &  $200$  & $(10,9,8)$   & $2.6$ & $550$ 	\\
R-G1S3    &    $(9,6,5)$ & $0.3$   &  $50$  & $(5,4,5)$ & $0.4$  &  $150$  & $(13,8,10)$ & $2.2$ & $750$ 	\\
R-G1S8    &    $(9,8,6)$  & $0.3$   &  $50$  & $(10,5,5)$  & $0.6$   &  $250$  & $(20,15,16)$ & $2.8$  & $1000$    	\\
R-G1S14    &    $(27,16,13)$  & $0.5$  &  $150$  & $(26,16,30)$   & $1.8$  &  $1200$  & $(30,23,28)$  & $3.4$ & $1700$ 	\\ \hline
\end{tabular}
\end{center}
\medskip
As in Table 4, symbols $*$ and $\dag$ denote quantities at the centre
and $4\,R_D$, respectively. 
\end{table*}

Due to this confusion about the role played by the global response of
the halo in the sinking process of the satellite we have repeated four
of our simulations (models G1S1, G1S3, G1S8 and G1S14) but replacing the
self-consistent halo by a rigid one which is free to move. With these new
simulations we try to cover several situations (different
eccentricities and satellite masses) to see whether the expression
(6) deviates from the results in our self-consistent N-body
calculations and from a satellite that does not follow a circular
orbit as is claimed by S\'eguin and Dupraz (1994, 1996). Figure 13
shows the sinking and disruption times for the satellites in the new
simulations. Open squares represent the fully self-consistent
simulations while solid lines correspond to the model with rigid
haloes. To quantify the sinking and distruption rates we have used
equation (6) with $\rho_H(r)=\rho(r)$ representing the local halo
density at the satellite position and the satellite mass $M_S=M_S(t)$
as the particles that remain bound. ${\bf F}_{df}$ is applied to all
the bound particles. Notice that the agreement is quite
remarkable and it suggests that a local description, as is implicit
in Chandrasekhar's formula, is adequate to determine the sinking
times of satellites whenever the satellite is immersed in the halo provided 
we choose the right value for $\Lambda $ and we define the satellite
mass $M_S$ as the total mass of bound particles. The values for
$\Lambda $ that appear in figure 13 were estimated initially from the 
expression $\Lambda=p_{max}/p_{min}$ \footnote{Here, $p_{max}$ and
$p_{min}$ are the maximum and minimum impact parameter, respectively,
and are defined by $p_{max}\equiv <R> \simeq GM^2/(2|W|)$ and
$p_{min}\equiv GM_S/\sigma_H^2(r_p)$ where $M$, $|W|$, $M_S$, 
$\sigma_H(r_p)$ are the total mass of the spiral galaxy, the total
potential energy of the spiral, the satellite's initial mass and the
one-dimensional dispersion velocity of the halo evaluated at the
pericentric radius $r_p$, respectively.} and afterwards were
fine-tuned to get the `right' values. We must point out that to
determine the dependence of $\Lambda$ on the orbital parameters and
the satellite structure further work is necessary.\\


\begin{figure}
\plotone{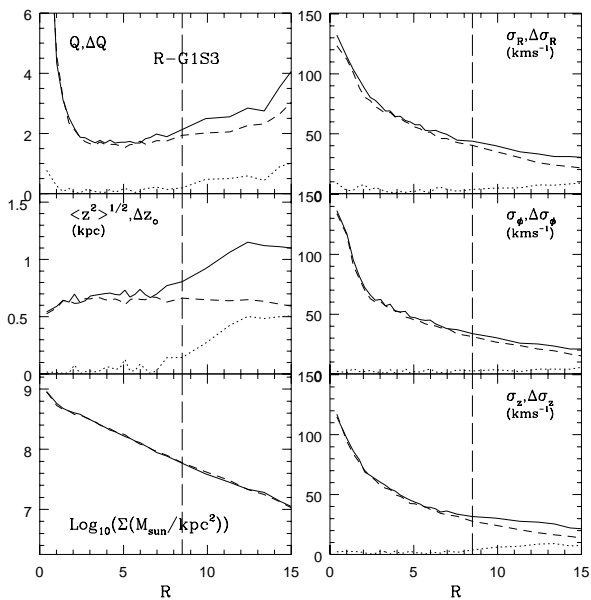}
\caption{Disc kinematics for model R-GS3 (solid
lines). Dashed lines represent the kinematics of the equivalent
isolated galaxy model while dotted lines correspond to the changes
suffered by the disc as a consequence of the satellite accretion.}
\label{Fig. 14}
\end{figure}

\section{ Rigid Haloes and Disc Heating and Thickening}

We have carried out a detailed analysis of the simulations involving rigid
haloes in order to address the importance of the
self-consistent 
representation of the haloes in the accretion of a satellite. In
figure 14 we show the resulting disc kinematics (solid lines) for
the case equivalent to our model G1S3 but with a rigid halo
(model R-G1S3). The kinematics of the corresponding isolated control
model is displayed in the same figure
(dashed lines) and the kinematical changes induced by the accretion
event are indicated by dotted lines. From comparison of figures 5 and
14 we can notice that: (i) the disc thickening due purely to numerical
relaxation has been diminished by the introduction of a rigid halo;
in an isolated model the disc heating is dominated by encounters
between disc and halo particles. (ii)
After completion of the accretion event, the radial and azimuthal
changes of the velocity ellipsoid induced by the satellite are
roughly similar to those found in model G1S3. Thus, limits on
accreted mass based only on the value of $\Delta Q$ ($\propto \Delta
\sigma_R$) and ignoring any change of $\Sigma_D(R)$ and any gas
cooling will be essentially the same even if the spherical halo is  
represented by a rigid potential rather than by a distribution of
particles provided the decay and disruption rates of the satellite are
considered adequately (in our case by equation 6). (iii) However, the
response of the vertical structure
of the disc seems to be strongly coupled to the responsiveness of the
halo. For example, for
model R-G1S3 we found that the change of the vertical scale length at
$14$ kpc is about a factor of $2$ bigger than in model G1S3 
while at the solar radius the factor is $\sim 1.5$. This difference is
substantial despite the fact that we have mimicked the satellite's
orbital energy loss to the rigid halo through equation 6. This may be
because the disc develops some other
instabilities, as satellite accretion proceeds, which can be damped
by a self-gravitating halo. (iv) As the
satellite becomes more massive, the necessity of a self-consistent
treatment of the halo is more evident; see figure 15
which is the equivalent of our model G1S14 (compare it with figure
6). The kinematics of this set of simulations using 
rigid haloes is summarized in Table 6.\\ 


\begin{figure}
\plotone{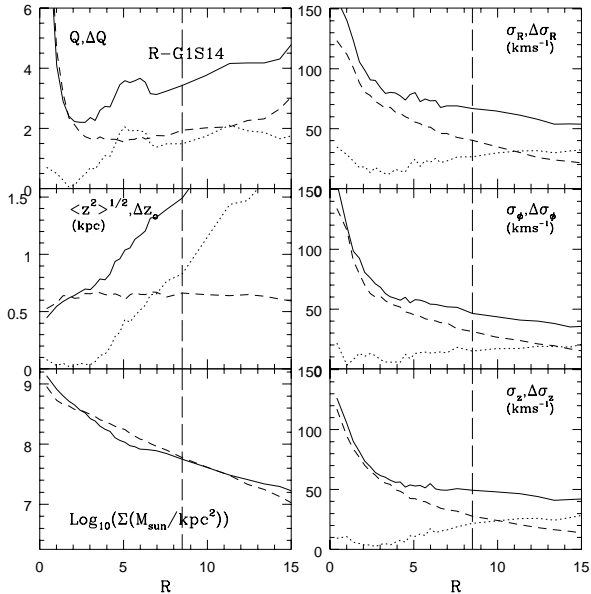}
\caption{The same as figure 14 but for a satellite 
with $0.2$ M$_D$. }
\label{Fig. 15}
\end{figure}

\section{ Conclusions}

We can summarize our main conclusions as follows:

\begin{itemize}
\item A comparison of our results with those obtained using TO's
formulae shows that the mass limits they derived are too strict. In
general, TO tend to
overestimate the damaging effects of satellite accretion by a
factor of about $2-3$ at the Solar radius. The
damaging effects in our simulations may also be an overestimation since
we ignore any contribution of gas cooling during the accretion process. The
origin of the discrepancy between TO's predictions and our results may
lie in the fact that: (i) their analysis ignores the coherent response of
the disc and its interaction with the halo. (ii) Their assumption that
the satellite's orbital
energy is deposited locally in the disc is clearly unrealistic. (iii)
A fully self-consistent treatment of the dynamics is needed to get reliable
results. A rigid halo (with dynamical friction introduced through
equation (6) leads to a larger increase of the vertical scale length of
the disc by a factor of $1.5-2$. A self-consistent treatment is more
important for more massive satellites.

\item The damaging
effects (heating and thickening) on the disc are very different for 
satellites on prograde and
retrograde orbits. A satellite on a prograde orbit tends to heat the
stellar disc while its retrograde counterpart induces a coherent
response of the disc in a form of a tilt. A massive satellite on a 
retrograde orbit may be accreted by a spiral galaxy without destroying its
disc. Furthermore, a massive central bulge may reduce the vertical heating 
of the disc for prograde orbits (but weaker for retrograde) 
while it may slightly diminish the tilting of the disc for retrograde 
ones (but not for prograde).

\item A satellite as massive as $0.1\,M_D$ and moving in a roughly
elongated orbit (e.g. our model G1S3) does not produce a
strong stellar warp. The
most noticeable case of warp formation occurs in our model G1S14 where
we see a departure from the disc plane of less than $7^o$ in the outer
regions (at about $15$ kpc). 

\item Chandrasekhar's dynamical 
friction formula gives remarkably good estimates for the sinking and
disruption rates of satellites in a variety of situations provided
a suitable value is chosen for the Coulomb logarithm and the
satellite mass is taken to be the mass still bound to
the satellite at each moment.

\end{itemize}

\section{ACKNOWLEDGMENTS}

We thank L. Hernquist for provide us with his algorithms. We also thanks 
an anonymous referee for his helpful comments to improve this paper. HV 
acknowledges useful conversations with L. Aguilar and S. Levine. HV 
thanks CONACyT (Consejo Nacional de Ciencia y Tecnolog\'{\i}a) of M\'exico 
for a studentship and the Max-Planck f\"ur Astrophysik for a Stipendium for 
the fulfillment of this research. Most of the simulations were run at the 
Max-Planck Society's Computer Centre at Garching, Germany. 


\end{document}